\documentclass[sigconf, nonacm]{acmart}

\AtBeginDocument{%
  }

\setcopyright{acmlicensed}
\copyrightyear{2018}
\acmYear{2018}
\acmDOI{XXXXXXX.XXXXXXX}
\acmConference[Conference acronym 'XX]{Make sure to enter the correct
  conference title from your rights confirmation email}{June 03--05,
  2018}{Woodstock, NY}
\acmISBN{978-1-4503-XXXX-X/2018/06}

\begin{document}

\title[Interaction-Driven Browsing]{Interaction-Driven Browsing: A Human-in-the-Loop Conceptual Framework Informed by Human Web Browsing for Browser-Using Agents}

\author{Hyeonggeun Yun}
\email{hg.russ.yun@companoid.io}
\orcid{0000-0002-9593-8282}
\affiliation{%
  \institution{Companoid Labs and Herbert Computer, Inc.}
  \city{Seoul}
  \country{Republic of Korea}
}

\author{Jinkyu Jang}
\email{jk.chairman@companoid.io}
\orcid{0000-0003-3150-0422}
\affiliation{%
  \institution{Companoid Labs and Herbert Computer, Inc.}
  \city{Seoul}
  \country{Republic of Korea}
}

\renewcommand{\shortauthors}{Yun et al.}

\begin{abstract}
Although browser-using agents (BUAs) show promise for web tasks and automation, most BUAs terminate after executing a single instruction, failing to support users' complex, nonlinear browsing with ambiguous goals, iterative decision-making, and changing contexts. We present a human-in-the-loop (HITL) conceptual framework informed by theories of human web browsing behavior. The framework centers on an iterative loop in which the BUA proactively proposes next actions and the user steers the browsing process through feedback. It also distinguishes between exploration and exploitation actions, enabling users to control the breadth and depth of their browsing. Consequently, the framework aims to reduce users’ physical and cognitive effort while preserving users’ traditional browsing mental model and supporting users in achieving satisfactory outcomes. We illustrate how the framework operates with hypothetical use cases and discuss the shift from manual browsing to interaction-driven browsing. We contribute a theoretically informed conceptual framework for BUAs.
\end{abstract}


\begin{teaserfigure}
  \centering
  \includegraphics[width=0.85\textwidth]{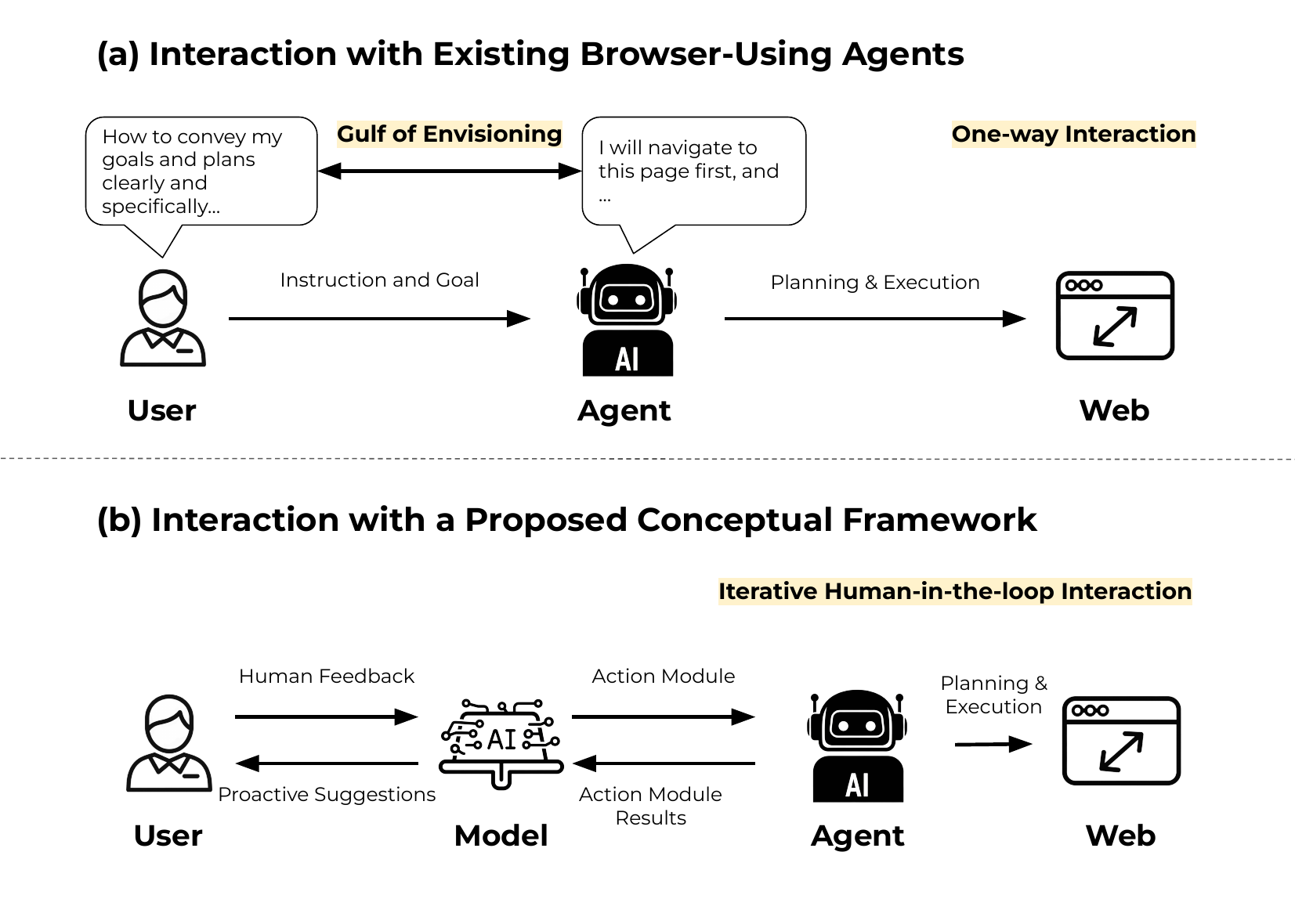}
  \caption{Comparison of existing browser-using agents (BUAs) and our proposed framework. (a) Interactions with existing BUAs present two major problems: (1) Gulf of Envisioning, where users must provide clear and specific instructions and plans to ensure the agent's actions align with their expectations, and (2) One-way Interaction, where most agents terminate after a single command, failing to maintain user context and previous action results. (b) Our proposed framework addresses these problems through an iterative human-in-the-loop (HITL) interaction. In this framework, proactive suggestions and human feedback bridge the gulf of envisioning, and the model generates action modules based on the feedback and maintains their results and the user's context throughout the loop.}
  \Description{
  Figure 1 is a diagram that compares two models of interaction with browser-using agents. It is divided into two horizontal panels, labeled (a) and (b). Part (a), titled "Interaction with Existing Browser-Using Agents," depicts a linear, one-way process. On the left, an icon of a "User" sends an arrow labeled "Instruction and Goal" to a central robot icon labeled "Agent." The Agent then sends an arrow labeled "Planning and Execution on Web" to a browser window icon on the right, labeled "Web." The entire process is described with a "One-way Interaction" label in the top-right corner. Above the flow, the concept "Gulf of Envisioning" is illustrated with a two-way arrow connecting two speech bubbles. The bubble originating from the user says, "How to convey my goal clearly and specifically...". The bubble originating from the agent says, "I will navigate to this page first, and...". Part (b), titled "Interaction with Proposed Conceptual Framework," depicts an iterative, multi-actor process. Introduces a new "Model" icon between the "User" and the "Agent." A two-way feedback loop exists between the User and the Model. An arrow from the User to the Model is labeled "Human Feedback." An arrow from the model back to the user is labeled "Proactive Suggestions." A second two-way flow exists between the Model and the Agent. An arrow from the Model to the Agent is labeled "Action Module," and a return arrow from the Agent to the Model is labeled "Action Module Results." Finally, a one-way arrow from the agent points to the "Web" icon, labeled "Planning and Execution."
  }
  \label{fig:teaser}
\end{teaserfigure}


\maketitle

\section{Introduction}

Recent and rapid advancements in (multi-modal) large language models ((M)LLMs) have stimulated interest and research in the field of artificial intelligence (AI) agents \cite{wang2024survey, hu-etal-2025-os, barua2024exploring}. This progress has led to the emergence of computer-using agents (CUAs), which assist or perform tasks for users in computer-based environments \cite{qin2024tool, liu2023visual, tan2025cradle, agashe2025agent2, kim2023language, sager2025comprehensive, hu-etal-2025-os, wu2024atlas, lin2025showui, liu2024autoglm, wu2024copilot, openai2025computer, anthropic2024introducing, wang2025opencuaopenfoundationscomputeruse, qin2025uitarspioneeringautomatedgui}, and browser-using agents (BUAs), also called web agents, which assist or perform tasks for users within web browsers \cite{gur2024real, agashe2025agent, iong2024openwebagent, abuelsaad2024agent, mozannar2025magentic, zheng2024gpt, gur2023understanding, furuta2024multimodal, lai2024autowebglm, he2024webvoyager, cai2025large, openai2025introducing, verma2024adaptagentadaptingmultimodalweb}. Due to their practical and potential applications, these agents are receiving significant attention from both academia and industry. Considering that most digital activities today are web-based, BUAs have the potential to improve user productivity in a wide range of tasks, including information retrieval, online shopping, document work, and automation of complex processes \cite{drouin2024workarena, ning2025survey, yao2022webshop, yoran2024assistantbench, peeters2025webmallmultishopbenchmark}. Operating in environments more constrained than CUAs, BUAs can achieve higher performance than CUAs utilizing the structured nature of the web \cite{abhyankar2025osworld, chi2024impactelementorderinglm}. This allows them to streamline complex processes and reduce user effort on the web.

However, current research on BUAs fails to adequately reflect the complexity of real-world user scenarios \cite{lù2025buildwebagentsagents, ye2025realwebassistbenchmarklonghorizonweb, yoran2024assistantbench, he2024webvoyager, verma2024adaptagentadaptingmultimodalweb}. Existing studies focus mainly on BUAs that execute a sequence of planned actions to achieve a clear and simple goal \cite{zheng2024gpt, gur2023understanding, furuta2024multimodal, abuelsaad2024agent, iong2024openwebagent, lai2024autowebglm, he2024webvoyager}. Many benchmarks and demonstrations assume simplistic scenarios in which BUAs complete a user's instruction within a single execution loop and then terminate \cite{zhou2024webarena, deng2023mind2web, shi2017world, drouin2024workarena, yao2022webshop, koh2024visualwebarena}. Generally, they concentrate on linear and structured tasks, such as searching for specific keywords and answering questions based on search results. In contrast, user browsing on the web is far more complex and nonlinear \cite{son2024unveiling, pirolli1995information, pirolli1999information, pirolli2003snif, pirolli2005sensemaking, white2009exploratory, marchionini2006exploratory, bates1989design}. Users often begin browsing with ambiguous or uncertain goals, dynamically setting multiple subgoals to achieve them. They engage in complex cognitive processes of decision-making, evaluating the outcomes of browsing at every moment and determining subsequent actions under uncertainty \cite{pirolli1995information, pirolli1999information, pirolli2003snif, pirolli2005sensemaking, white2009exploratory, marchionini2006exploratory, bates1989design}. This natural human approach to browsing is an iterative process of reaching goals through multiple stages of actions, rather than executing a single instruction at a time.

This gap often results in BUA behavior resembling routine automation, which limits performance on complex and unstructured goals in web environments. To achieve complex goals through a human-like decision-making process and translate them into effective browsing actions, a human-in-the-loop (HITL) process is essential \cite{mosqueira2023human, amershi2014power, mozannar2025magentic, huq2025cowpilot, cheng2025navi, akbar2024towards, natarajan2025human}. The process involves user feedback, which consists of evaluating the BUA’s results and actively determining the direction of the next action. User feedback is particularly important when the BUA encounters unexpected situations, needs to interpret ambiguous goals, or when user preferences and intentions change \cite{cheng2025navi}. However, from an HITL perspective, research on BUAs has been limited, particularly regarding how users can provide real-time feedback to guide subsequent actions of BUAs \cite{mozannar2025magentic, huq2025cowpilot}. Existing research tends to focus on agent autonomy or utilizes user feedback only in the pre-planning or post-verification stages for privacy and security issues, thus limiting the potential for adaptive decision-making on the browsing direction through an HITL interaction.

To overcome the limitations, this paper proposes a conceptual framework for the HITL interaction designed to effectively support complex web browsing with BUAs. This framework establishes an iterative feedback loop between the user and the BUA to achieve a user's ambiguous and multifaceted goal. We apply principles and theories of human behavior in web browsing to the concept of the framework. Therefore, the framework naturally integrates and reflects the human web browsing process, including the depth of browsing, the timing of switching browsing targets, and the decision-making processes regarding information exploration and exploitation. This allows users to perform browsing in a manner similar to human browsing, solely through interacting with BUAs. Furthermore, the roles of the user and the BUA are clearly distinguished: the user focuses on decision-making and evaluating browsing results, while the BUA concentrates on performing actions on the web pages. As a result, the user may shift physical and cognitive effort away from low-level interactions and focus cognitive effort on higher-level decision-making over the entire browsing process, thereby conducting it as desired.

The contribution of this paper is primarily theoretical and conceptual. Rather than implementing a system and presenting an empirical evaluation, we aim to provide a new theoretical lens for understanding and designing the next generation of browsing through agents. Therefore, this paper introduces theories from HCI and cognitive science related to web browsing and, based on these theories, presents a novel conceptual framework for HITL interaction with a BUA in web browsing. The framework offers mechanisms and concepts for supporting user decision-making processes stemming from ambiguous and unclear goals, and for conducting the complex, nonlinear browsing characteristic of human behavior. Subsequently, we use hypothetical user scenarios to illustrate how the framework can perform and support browsing according to the user's goals. Finally, we discuss the potential for \textbf{interaction-driven browsing} with a BUA to become the next paradigm in browsing and present the design and theoretical implications of our proposed framework.

\section{Background and Related Work}
\subsection{Browser-Using Agents and Their Challenges}\label{background-bua}
Recent advancements in the image understanding, coding capabilities, and tool-calling functions of (M)LLMs have spurred research on computer-using agents (CUAs) that can manipulate computers according to user requests \cite{qin2024tool, liu2023visual, tan2025cradle, wang2024survey, hu-etal-2025-os, kim2023language, sager2025comprehensive, hu-etal-2025-os, wu2024atlas, lin2025showui, liu2024autoglm, wu2024copilot, wang2025opencuaopenfoundationscomputeruse, agashe2025agent2, openai2025computer, anthropic2024introducing, qin2025uitarspioneeringautomatedgui}. Among these, CUAs that focus on browsing within a web browser are called browser-using agents (BUAs) \cite{gur2024real, iong2024openwebagent, abuelsaad2024agent, mozannar2025magentic, furuta2024multimodal, gur2023understanding, zheng2024gpt, shen2024scribeagentspecializedwebagents, lai2024autowebglm, he2024webvoyager, cai2025large, verma2024adaptagentadaptingmultimodalweb, openai2025introducing}. The browser, which is a more constrained environment than a computer operating system, provides not only visual cues based on images but also textual context that helps agents understand the environment, such as a HTML-based document object model (DOM) interface and accessibility trees \cite{abhyankar2025osworld, chi2024impactelementorderinglm, gur2023understanding}. Therefore, BUAs have a high potential for performance improvement and applicability for end-users. As a result, there has been a recent surge in commercial products, such as OpenAI's Operator \cite{openai2025introducing, anthropic2024introducing}, as well as various open-source projects \cite{browser_use2024, eko2025} and research \cite{gur2024real, iong2024openwebagent, abuelsaad2024agent, mozannar2025magentic, furuta2024multimodal, gur2023understanding, zheng2024gpt}.

Research on BUAs has concentrated on foundation models and agent frameworks to enhance an agent’s ability to understand and perform the actions required for user-requested browsing tasks \cite{zheng2024gpt, abuelsaad2024agent, iong2024openwebagent, gur2024real, shen2024scribeagentspecializedwebagents, lai2024autowebglm, cai2025large, verma2024adaptagentadaptingmultimodalweb}. Zheng et al. demonstrated the potential of BUAs by combining the visual understanding capabilities of GPT-4V with actions such as clicking and typing \cite{zheng2024gpt}. However, they pointed out that the grounding process, accurately connecting model-generated text and plans to actions on actual web elements, remains a significant limitation. In contrast, some research, such as Agent-E, has utilized the DOM interface instead of images, achieving a high success rate of 73.2\% on the WebVoyager benchmark \cite{abhyankar2025osworld}. Shen et al. proposed ScribeAgent, which was developed by collecting data from real user tasks on various websites and fine-tuning an open-source LLM \cite{shen2024scribeagentspecializedwebagents}. Despite having fewer parameters than general models, ScribeAgent outperformed GPT-4-based agents on the WebArena and Mind2Web benchmarks.

More recently, research has expanded from foundation models and agent frameworks to training frameworks \cite{zhang2025symbioticcooperationwebagents, hongjin2025learn}. Zhang et al. proposed the AgentSymbiotic framework, where an LLM generates high-quality task trajectories, and a small language model (SLM) explores new paths \cite{zhang2025symbioticcooperationwebagents}. These models share data, complementing each other to enhance performance. Similarly, Su et al. introduced the Learn-by-Interact framework, which enables an agent to automatically generate and learn high-quality data without human labeling, leading to significant performance improvements on the WebArena benchmark \cite{hongjin2025learn}.

Although research on BUAs has primarily focused on task execution capabilities and model training methods, there are several limitations in their practical use by end-users (Figure~\ref{fig:teaser}). Current benchmarks often use simulated websites and do not adequately cover complex real-world browsing tasks such as navigating multiple websites, transitioning between several tabs, or integrating various information from web pages \cite{zhou2024webarena, yao2022webshop, deng2023mind2web, ye2025realwebassistbenchmarklonghorizonweb, yoran2024assistantbench, shi2017world, lu2024weblinx}. Furthermore, users may experience the gulf of envisioning when using BUAs \cite{subramonyam2024bridging, son2024unveiling, he2025plan, cheng2025navi}. Users tend to give ambiguous, high-level goals or instructions rather than specific plans, yet they expect the outcomes to align with their mental model. In this case, the BUAs make and execute specific plans themselves, but the plans and results are likely to diverge from the users' expectations. Moreover, in contrast to human adaptability, BUAs struggle to flexibly handle unexpected situations like broken links or failed searches, because it is impossible to account for all variables of complex web environments in the initial planning stage \cite{son2024unveiling, he2025plan, yao2023react, shlomov2024groundingplanningbenchmarkingbottlenecks, cheng2025navi}. This problem is exacerbated by the one-way interaction model adopted by most BUAs; they execute a single command and terminate, lacking the ability to maintain context across multiple tasks or handle failures appropriately.

To address these issues, an HITL approach is needed, allowing users to clarify goals and provide context through interaction before and after the execution of a BUA. Although some BUA studies have incorporated HITL \cite{mozannar2025magentic, huq2025cowpilot, cheng2025navi}, their application have focused on resolving security and privacy issues during browsing, rather than bridging the gulf of envisioning. This paper proposes a conceptual framework based on the HITL interaction to overcome the limitations of the gulf of envisioning and one-way interaction.

\subsection{Theoretical Background of Web Browsing Behavior}\label{background-theory}
Web browsing is a complex decision-making process in which users explore, compare, and select information from various web pages to achieve a specific goal \cite{marchionini2006exploratory}. Therefore, understanding a user's browsing behavior requires an understanding of the cognitive theories related to information seeking and human decision-making.

Herbert Simon's theory of bounded rationality explained that humans are not perfectly rational in decision-making \cite{simon1972theories, simon2018satisficing, schwarz2022bounded}. According to the theory, humans tend to choose good enough solutions within the constraints of limited knowledge, time, and cognitive resources, rather than making optimal decisions. This is known as a strategy named ``satisficing.'' This tendency is clearly demonstrated in web browsing \cite{mansourian2007search, agosto2002bounded}. For instance, due to various constraints, it is difficult for users to explore every link in search results. Instead of checking every page of the search results, users might look at the top few pages or gather a moderate amount of information by switching between multiple tabs. Accordingly, they internally set a satisfactory standard and continue browsing until the standard is met. This implies that browsing outcomes can vary depending on the user's browsing skills and criteria.

In pursuit of satisfactory outcomes, users constantly face the exploration-exploitation dilemma, also known as the exploration-exploitation trade-off, deciding at every moment whether to explore more broadly or to exploit the information collected so far \cite{berger2014exploration, cohen2007should}. In web browsing, exploration refers to navigational acts to find new information, such as searching a new query or moving to a new web page. In contrast, exploitation implies using and consuming information that has already been discovered and collected, such as evaluating and comparing various information or making a decision based on that comparison. During browsing, users decide whether to explore new information to find better alternatives or to exploit options from the information they have explored so far \cite{wiebringhaus1measuring}. For example, a user trying to buy the lowest price product on an online shopping website explores various products and websites \cite{browne2005stopping}. At some point, they must decide whether to exploit the best option found so far or to continue exploring for a better one. This dilemma arises because users cannot search all information on the web due to time and cognitive limitations, forcing them to constantly compare the uncertain value of new information against the certain utility of information already found.

Information foraging theory (IFT), proposed by Pirolli and Card, offers a useful model from a cost-reward perspective for how users resolve this dilemma \cite{pirolli1995information, pirolli1999information}. IFT analogizes web browsing to animal foraging behavior. In ecology, optimal foraging theory posits that animals use strategies to gain the maximum benefit (food) for the minimum cost (energy and time). Similarly, IFT suggests that users browse in a way that yields the most valuable information for the least amount of time and cognitive effort. Users navigate based on information scent, cues such as link titles, summaries, or URLs that suggest the likelihood of finding the desired information on a page. A collection of websites or pages with a strong information scent forms an information patch. A user exploits information within one patch until the information scent weakens. At that point, they may decide that the benefit of moving to a new patch outweighs the cost, and resume exploration by trying a different search query. Therefore, according to the IFT framework, users browse with a strategy of seeking the most valuable information while minimizing resource costs based on information scent, repeating this process until they have gathered a satisfactory level of information.

In summary, human web browsing is an act of seeking satisfactory information and results within limited resources, driven by decision-making within the exploration-exploitation dilemma. Therefore, BUAs must support the significant cognitive and temporal resources used in the user's browsing process, reflecting users' browsing patterns to achieve a sufficiently satisfactory level of web browsing. However, current BUAs only support a small part of the browsing process and do not support the decision-making process of users. This paper, therefore, proposes a framework based on the HITL interaction that reflects human web browsing patterns, suggesting ways for BUAs to assist in the information-seeking and decision-making processes of human browsing.

\subsection{Human-in-the-loop in AI and HCI}\label{background-hitl}
Human-in-the-loop (HITL) is a paradigm that integrates human interaction into the lifecycle of AI systems to improve their performance, safety, and alignment \cite{mosqueira2023human}. As machine learning (ML)-based AI systems cannot always guarantee consistent results and performance, research on HITL has been conducted to overcome these limitations. Representative HITL methodologies include the following: Active Learning \cite{settles2009active-key, olsson2009literature}, Interactive Machine Learning (IML) \cite{amershi2014power, fails2003interactive}, Machine Teaching \cite{simard2017machine}, and Reinforcement Learning from Human Feedback (RLHF) \cite{kaufmann2024surveyreinforcementlearninghuman, christiano2017deep}.

Active learning begins training with a small amount of labeled data and enhances training efficiency by having the model request human labeling for the data it deems most beneficial for training \cite{settles2009active-key, olsson2009literature}. Active learning is useful when labeling is costly or data is large, but it has several limitations such as human error and the risk that the model requests labels for incorrect data samples \cite{settles2009active-key, settles2011theories, donmez2008proactive, cakmak2010designing}. Interactive machine learning (IML) involves more direct and iterative user participation in the model training loop \cite{amershi2014power, fails2003interactive, porter2013interactive}. Users interact with the model in various ways beyond labeling, such as adjusting the training direction, providing feedback, and reviewing results to progressively improve the model. IML allows even non-experts to participate and quickly refine a model to reflect desired outputs \cite{michael2020interactive, amershi2014power, ware2001interactive, dudley2018review, fiebrink2011human, delcourt2024human}. However, the quality of the results depends on the user's feedback, and the process requires significant user time and effort. Furthermore, research on interfaces is also important in IML to help users provide feedback and intervene in the training loop \cite{fiebrink2011human, dudley2018review, fails2003interactive, Füßl2025design}.

Machine teaching is a process led by a domain expert who directly imparts important samples, rules, and knowledge to the model \cite{zhu2015machine, zhu2018overview, simard2017machine}. It is effective for building models with limited data by incorporating expert intuition, but it is demanding in terms of human effort required to design the entire training process and can be limited in its ability to handle complex problems \cite{simard2017machine, peng2021soloist, ramos2020interactive, devidze2021understanding}. Finally, widely used reinforcement learning from human feedback (RLHF) trains an agent by learning a reward model based on human preference feedback, instead of relying on a pre-designed reward function \cite{kaufmann2024surveyreinforcementlearninghuman, christiano2017deep}. This method is used in LLM fine-tuning, robotics, and gaming \cite{chen2023feedback, xu2023imagereward, lee2023aligningtexttoimagemodelsusing, ibarz2018reward, xu2020playing, hejna2023few}, and excels at aligning models with user goals and solving complex problems without clear reward functions \cite{ibarz2018reward, christiano2017deep, wirth2017survey, ziegler2019fine, ouyang2022training, bai2022traininghelpfulharmlessassistant}. However, collecting consistent and reliable human feedback is challenging, especially when user preferences are unclear \cite{lee2024rlaif, bai2022traininghelpfulharmlessassistant, ouyang2022training, bukharin2024robust, shi2025dxhf}.

These HITL methodologies are significant for addressing the limitations of ML-based AI systems through user interaction. However, there are several limitations, such as inconsistencies in feedback quality among users and the significant effort required from them. Consequently, the field of HCI has conducted research to support these HITL methods.

HCI research aims to alleviate the difficulties users face in HITL. For active learning, to reduce user fatigue, Cakmak and Thomaz proposed ``intermittently active learning,'' where the model requests labeling only at important moments or with explicit user permission \cite{cakmak2010designing}. For IML, new interfaces and tools have been developed to facilitate the participation of non-experts \cite{dudley2018review}. For example, the tool named Crayons allows users to provide feedback for image segmentation models by drawing lines on an image \cite{fails2003interactive}. Mishra and Rzeszotarski introduced a tool with a drag-and-drop interface that enables non-experts to conduct interactive transfer learning by combining convolutional neural network (CNN) model layers \cite{mishra2021designing}. Françoise et al. proposed the toolkit named Marcelle, which provides an environment for both non-expert users and expert users to participate and collaborate throughout the entire IML workflow, from data collection to model evaluation \cite{franccoise2021marcelle}.

In machine teaching, research has focused on enabling users to transfer knowledge effectively. MacLellan et al. proposed a human-centered framework for designing teachable systems where users can directly shape an AI's behavior \cite{maclellan2019human}. Furthermore, Zhou suggested methods for interactive machine teaching that use saliency map visualizations to help users understand model predictions, gestures to specify object locations, and interfaces that visualize data diversity in real-time to help users teach the model from multiple perspectives \cite{zhou2022exploiting, zhou2021enhancing, zhou2022gesture}. For RLHF, research has focused on interfaces that reduce cognitive load during the feedback process. Shi et al. proposed a decomposition-based interface that breaks long-form texts into atomic claims to compare LLM results, making it easier for users to identify key differences \cite{shi2025dxhf}.

As demonstrated, research on HITL in AI has addressed the limitations of ML models through human interaction, and this human-engaged loop has proven effective. In parallel, the field of HCI has developed interfaces and interactions to support these HITL-based systems. The synergy between AI and HCI has been crucial in advancing HITL systems, showing that overcoming AI's limitations requires not only better algorithms but also better human-AI interaction design. Similarly, as HITL research has improved existing AI fields, this paper aims to utilize the concept of HITL to design a framework that can overcome the limitations of BUAs.

\section{Conceptual Framework}
In this section, we propose a conceptual framework for the HITL interaction to address the problems of existing BUAs. First, we define the fundamental concepts and roles of actors that constitute the framework. Then, we describe design principles of the framework, based on theories related to human browsing behavior. Finally, we detail how these principles are integrated into the iterative HITL interaction flow.

\subsection{Fundamental Concepts}

\subsubsection{Framework Terminology}
The proposed framework uses four hierarchical concepts to execute a user's request to a BUA based on an HITL approach: Goal, Subgoal (Task), Action Module, and Action.

\begin{itemize}
    \item Goal: A goal is an ultimate objective the user aims to achieve through interaction with the BUA. It corresponds to the user's initial request and can be a simple goal consisting of a single subgoal (task) or a complex goal composed of several subgoals (tasks). For instance, it can be a simple goal such as \textit{``Buy milk''} or a complex goal such as \textit{``Find types of browser-using agents, write a report on them, and send it by email.''}
    \item Subgoal (Task): A subgoal, also called a task, is a detailed objective that must be accomplished to achieve the goal. A goal is naturally decomposed into subgoals (tasks). For a simple goal such as \textit{``Buy milk,''} the goal and the subgoal (task) are identical. A complex goal like \textit{``Find types of browser-using agents, write a report on them, and send it by email''} is broken down into multiple subgoals (tasks): \textit{``Find types of browser-using agents,''} \textit{``Write a report on the types of browser-using agents,''} and \textit{``Send the report by email.''}
    \item Action Module: An action module is a meaningful sequence of continuous actions performed by the BUA in the web environment to achieve a subgoal (task). It represents a small, independent sequence of actions completed within a specific website, domain, or service. For example, under the subgoal (task) \textit{``Find types of browser-using agents,''} various action modules could exist, such as \textit{``Search for browser-using agents on Google Scholar''} or \textit{``Find papers related to browser-using agents in the ACM Digital Library.''} The action modules are dynamically generated in each loop, reflecting user feedback, context, and the results of the previous action modules.
    \item Action: An action is the smallest unit of physical behavior that constitutes an action module. It refers to individual operations performed by the BUA, such as manipulating DOM interface elements, navigating to a URL, entering text, or clicking a button. For instance, the action module \textit{``Search for papers on Google Scholar''} consists of a series of specific actions such as \textit{``Navigate to URL scholar.google.com,''} \textit{``Click the search bar element,''} \textit{``Input text into the search bar,''} and \textit{``Click the search button.''}
\end{itemize}

\subsubsection{Role of Actors}
An actor is an entity that performs any work or operation such as input, decision-making, reasoning, or execution within the framework. The framework consists of three actors: User, Model, and Agent.

\begin{itemize}
    \item User: An user is the ultimate decision-maker in the framework. The user not only sets the goal, but also orchestrates and controls the direction of the entire browsing process. In each loop, the user judges the value of the information brought by the agent, injects their unique context (e.g., personal preferences, prior experience, hidden constraints) into the framework, and provides feedback that determines the direction of the next action module. As a result, the user plays a central role in supervising the agent's actions and leading the entire process to achieve their goal.
    \item Model: A model is the mediator between the user and the agent. This role is typically performed by a language model (LM). The model summarizes the results of the action modules executed by the agent and proactively suggests potential next action modules based on the previous results. Furthermore, it interprets user feedback and changes it into a specific structured action module that the agent can execute. In addition, it is responsible for maintaining results and context across multiple subgoals (tasks) and action modules.
    \item Agent: An agent is an executor of the actual web browsing. According to the action module received from the model, it executes specific actions such as navigating web pages and extracting information, and then reports the results to the model. This role is typically fulfilled by a BUA.
\end{itemize}

\subsection{Framework Design from Theoretical Foundation}
As described in Section 2, human web browsing behavior can be explained by several background theories. This framework is designed to support the complex physical and cognitive processes of how humans explore information and make decisions on the web through a BUA. Thus, we have designed the framework based on four theoretical foundations that reflect the human web browsing process.

\subsubsection{Bridging the Gulf of Envisioning}
The gulf of execution is a problem that arises from the discrepancy between a user's mental model and the intentions for a system and the actual actions they must take to operate it \cite{hutchins1986direct}. In LLM-based systems, this problem extends to the gulf of envisioning, where users struggle to know what prompt to input to get their desired results \cite{subramonyam2024bridging}. In the gulf of envisioning, the user only conveys their goals to the LLM via prompting, without specific plans or instructions to achieve the goals. As explained in section~\ref{background-bua}, the same problem occurs in the interaction with a BUA. For instance, a user might give a BUA a general and broad goal such as \textit{``Buy milk.''} However, in order to understand the hidden intent of the user and derive satisfactory results, it needs to know details such as which service to use, what kind of milk to buy, and what is important when purchasing milk. Otherwise, the BUA might operate in a way that contradicts the user's implicit intentions.

\paragraph{Reflected Design} 
To address the gulf of envisioning, our framework proposes two interaction concepts: Proactive Suggestions and Human Feedback.

\begin{itemize}
    \item Proactive Suggestions: Previous research has suggested that the LLM can propose next steps before the user's prompt or command since it is cognitively demanding for users to plan specific instructions \cite{lu2025proactive, chen2025need, jones2024designing}. Based on previous research, the model does not passively wait for explicit user requests or feedback in the framework. Instead, it proactively asks questions to clarify the user's ambiguous goal and context, or predicts and suggests several action modules based on the context and results from previous action modules. For example, if the goal is \textit{``Buy milk,''} proactive suggestions could provide questions to gather the necessary context, such as \textit{``What platform or service would you like to buy milk from?''} or \textit{``What kind of milk do you prefer?''} It could also suggest action modules based on previous results or context, such as \textit{``Search for fat-free milk on Amazon''} or \textit{``Compare the prices of the milk investigated so far.''} Proactive suggestions help users articulate their vague thoughts into specific requests and requirements, potentially reducing decision effort. Consequently, users can more easily convey their implicit intentions to the agent and achieve the browsing results they envisioned.
    \item Human Feedback: This is the process in which the user reviews the previous results and the model’s suggestions and conveys their own context or intent. Human feedback consists of two concepts: Context Injection and Decision-making.
    \begin{itemize}
      \item Context Injection: The user adds the necessary contextual information during the creation and execution of an action module to refine it. The user provides additional information necessary to generate a more accurate and effective action module, such as their intentions, preferences, or current situation. For example, for the goal of \textit{ ``Buy milk''}, they could provide information such as \textit{``Find it on Amazon''} or \textit{``The budget is \$10.''} This context helps the agent perform actions that align with the user's usual web browsing context and behavioral patterns.
      \item Decision-making: The user determines the direction of the action module. Based on the action modules suggested by the model, the user decides on the next action module. First, the user judges whether satisfactory information has been collected from the results provided by the model. If not, they can instruct for further exploration or request an exploitation of the collected information to analyze and integrate the results. For example, for the goal of \textit{``Buy milk''}, if the user wants to explore other options, the user might instruct, \textit{``Investigate other brands of milk.''} If the user wants to compare the results explored based on what is important (e.g. price), the user might decide, \textit{``Compare the prices of the milk explored so far.''} In short, through decision-making, the user proactively leads the type and direction of the action module and the entire goal-achievement process.
    \end{itemize}
\end{itemize}

\subsubsection{Information Foraging Theory}
As explained in section~\ref{background-theory}, IFT describes human information-seeking patterns as being similar to animal foraging strategies \cite{pirolli1995information, pirolli1999information}. Users select web pages (patches) based on the information scent and explore them. They repeatedly move between pages based on the perceived value of information, considering the trade-off between the potential gain from new information and the cost (time, cognitive load, etc.) of searching for it. This implies that humans search for information on the web by finding the most efficient path to explore information under limited resources.

\paragraph{Reflected Design}
The framework is designed to reflect this information seeking pattern while reducing the browsing costs and increasing the information gain. The trade-off between gain and cost in traditional browsing arises since the user must track information scent, navigate pages, find information, and evaluate its value alone. Our framework distributes these roles among the agent, model, and user. To reduce costs, the agent performs repetitive and tedious actions such as visiting websites, clicking links, and scraping information on behalf of the user. Therefore, the framework is able to reduce the physical, temporal, and cognitive burden. To increase gain, the model presents key information from the agent’s results of the action modules to the user and provides proactive suggestions to help the user decide the direction of the action module for browsing more valuable information based on current results and context. Therefore, the user can focus on evaluating the information value and deciding on the next action module. As a result, the \textit {user} can perform web browsing that improves the value of information while reducing the effort required to obtain it through interaction.

\subsubsection{Satisficing}
Satisficing is a decision-making strategy in which humans, in situations where optimal decision-making is difficult, explore various alternatives until an acceptable threshold is reached, leading to a good enough decision \cite{simon1972theories, simon2018satisficing, schwarz2022bounded}. As explained in section~\ref{background-theory}, this is because finding the optimal alternative is practically impossible due to limited physical, temporal, informational, and cognitive resources. Users also employ a satisficing strategy during web browsing. Since the amount of information on the web is huge, users cannot explore it all. Therefore, they set a threshold for what constitutes sufficient information to achieve their goal and continue or repeat their browsing until the threshold is reached. During the browsing process, they evaluate the value of information and continue exploring until the quantity and value of information reach their personal threshold, at which point they stop.

\paragraph{Reflected Design}
The framework reflects the satisficing strategy. It is designed to allow users to repeat browsing through an iterative HITL interaction until they gather and obtain a level of information that is good enough. Within the loop, the user has complete decision-making authority to judge whether satisfactory information has been collected based on the results of action modules. Based on the judgment, the user also has complete decision-making authority to either continue or stop the iterative HITL interaction. This allows the user to direct the model and the agent to generate and execute action modules repeatedly until their threshold is met. In particular, since the framework reduces the costs related to physical, temporal, and cognitive burdens, users are more likely to set a higher threshold and reach it relatively easily.

\subsubsection{Exploration and Exploitation}
The exploration and exploitation trade-off is a fundamental concept in decision-making, referring to the choice a user makes between exploring new possibilities to find better results or exploiting the best choice within the information collected so far \cite{berger2014exploration, cohen2007should}. As explained in section~\ref{background-theory}, this also occurs in web browsing, where users decide at every moment whether to visit new web pages to explore new information or to exploit the best information based on what they have found so far.

\paragraph{Reflected Design}
To reflect the user's exploration and exploitation trade-off in the browsing process, the framework categorizes action modules into two types: Exploration Action Modules and Exploitation Action Modules. This allows the user to control whether the agent should explore new information on the web, such as by searching and navigating, or exploit previously collected information, such as by comparing and evaluating it.

\begin{itemize}
    \item Exploration Action Module: The exploration action module includes all action modules in which the agent collects new information or inputs data through direct interaction with the web. For instance, in the \textit{``Buy milk''} goal, an exploration action module could range from searching actions such as \textit{``Search for milk on Amazon''} and \textit{``Check details of a specific milk product''} to data entry actions required to achieve the goal, such as \textit{``Enter shipping information to purchase the milk product.''} Typically, there is at least one exploration action module within a single subgoal (task), and multiple may exist depending on the scope and depth of browsing. Since exploration action modules are generated and executed based on human feedback, the number and execution method of exploration action modules can vary based on the user's requested scope and depth of exploration, satisfaction with the explored information, and task complexity.
    \item Exploitation Action Module: The exploitation action module refers to all action modules that derive meaningful results by comparing, summarizing, analyzing, evaluating, or reasoning on the information collected through various exploration action modules, without new web exploration and interaction. For example, in the goal of \textit{``Buy milk''}, an exploitation action module could be \textit{``Compare the prices of the milk products found so far''} or \textit{``Analyze the ingredients of the milk products found so far.''} The results of an exploitation action module can help the user decide the direction of future action modules. For instance, if the cheapest product is found through the \textit{``Compare prices''} action module, the user might request an action module to purchase the cheapest product, thus shifting the focus of the browsing phase from searching to purchasing. If the results of the exploitation action module are not yet acceptable, the user can request additional information exploration by requesting another exploration action module and then attempt to derive meaningful results to achieve the goal again through a subsequent exploitation action module.
\end{itemize}

The exploration and exploitation action modules appear iteratively in the HITL interaction. Based on the information collected through the exploration action modules, the user can be assisted in deciding the next browsing action through the exploitation action modules, and then execute the decided action through another exploration action module. In this way, the user can perform browsing similar to traditional manual browsing by repeatedly generating and executing exploration and exploitation action modules through interaction with the model and the agent.

\subsection{Iterative Human-in-the-Loop Interaction Flow}

\begin{figure*}[h]
  \centering
  \includegraphics[width=\linewidth]{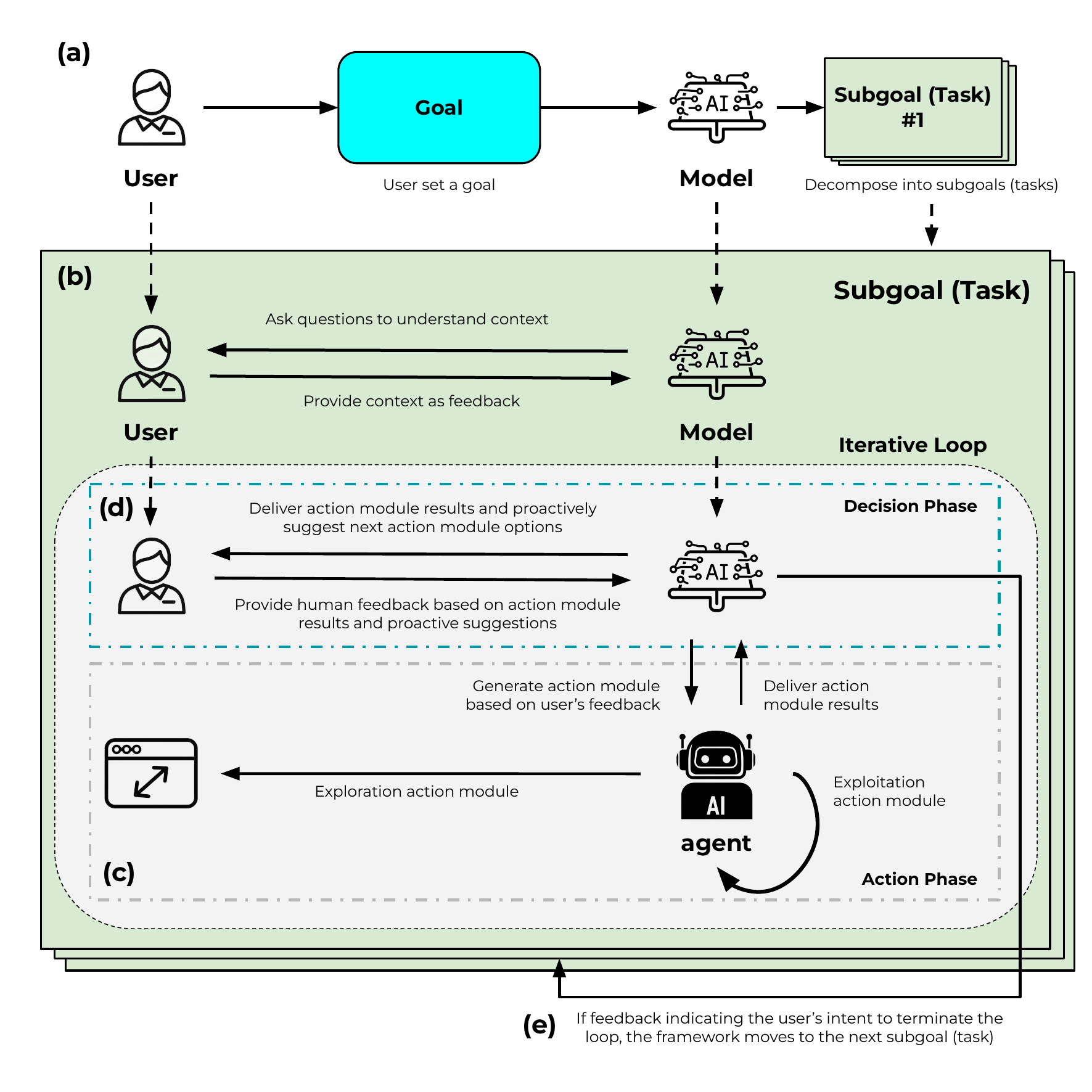}
 \caption{The detailed interaction flow of the proposed human-in-the-loop (HITL) conceptual framework. (a) User sets a high-level ambiguous goal, which the model decomposes into subgoals (tasks). (b) For each subgoal (task), the model first asks questions to gather context and user provides context as feedback. (c, d) The core of the framework is an iterative HITL loop, which consists of an action phase and a decision phase. (c) In action phase, an action module (either for exploration or exploitation) is generated based on the user's feedback and executed by the agent on the web. (d) In decision phase, the model delivers the results of an action module and proactively suggests next action modules, and the user provides feedback to guide the entire browsing process. (e) This loop continues until the user's feedback indicates an intent to terminate, at which point the framework moves to the next subgoal (task). }
  \Description{
  Figure 2 is a diagram detailing the interaction flow of the proposed human-in-the-loop (HITL) framework. The process is divided into five stages, labeled (a) through (e), progressing from top to bottom. Part (a) shows the initial Goal Setting and Decomposition: A 'User' icon points to a 'Goal' box, which in turn points to a 'Model' icon that decomposes the goal into 'Subgoal (Task) #1'. Part (b) depicts the Initial Context Injection within a larger container for the Subgoal. It shows a two-way interaction where the 'Model' asks questions to understand context and the 'User' provides context as feedback. Parts (c) and (d) illustrate the core 'Iterative Loop', which is divided into a 'Decision Phase' and an 'Action Phase'. The Decision Phase (d) shows a two-way interaction where the 'Model' delivers results and proactively suggests options to the 'User', who then provides human feedback. The Action Phase (c) shows an 'agent' icon executing an action module. An arrow for an 'Exploration action module' points to a browser window, while a looping arrow for an 'Exploitation action module' points back to the agent. The agent then delivers the results back up to the Model in the Decision Phase. Part (e) indicates the Termination Condition with an arrow exiting the iterative loop, connected to a text label explaining that the loop terminates based on the user's intent, at which point the framework moves to the next subgoal.
  }
  \label{fig:flow}
\end{figure*}

The core concept of the framework lies in the iterative HITL interaction, which reflects the user's individual browsing style and allows browsing until they are sufficiently satisfied with the current browsing results. During the loop, instead of manually navigating through tabs and windows in the browser, the user interacts with the model and the agent for web browsing. This \textbf{interaction-driven browsing} idea is achieved through the previously designed concepts of proactive suggestions, human feedback, and exploration and exploitation action modules.

When the user inputs a goal, it is decomposed into subgoals (tasks). Within each subgoal (task), the HITL interaction repeats until the purpose of the subgoal (task) is achieved. In the interaction, the user provides feedback based on the results of previous action modules, the model generates an action module that reflects the context and feedback of the user, and the agent executes the action module. The user continues the interaction until the results are satisfactory. The end of the loop indicates the completion of the subgoal (task), and the process moves on the next subgoal (task). The results from previously completed subgoals (tasks) can be utilized in subsequent subgoals (tasks).

Within the iterative HITL interaction, the phase in which the agent executes an action module and reports the results to the model is called an action phase. The phase between action modules, where the model presents results and proactive suggestions to the user and the user provides human feedback, is called a decision phase. Thus, a single interaction loop contains both an action phase and a decision phase, and the repetition of these two phases constitutes the flow of the interaction loop (Figure~\ref{fig:flow}). The flow of the framework is as follows.

\newcommand{\stagetag}[1]{\noindent\\\emph{[#1]}\\}

\begin{enumerate}
  \item \textbf{Goal Setting and Subgoal (Task) Decomposition}\\
  The user sets an overall goal. The model decomposes this goal into one or more subgoals (tasks) to establish clear purposes.

  \stagetag{Subgoal (Task) Start}

  \item \textbf{Initial Context Injection for the Subgoal (Task)}\\
  The model asks for the initial context required to generate the first action module within the first subgoal (task), and the user provides feedback.

  \stagetag{Iterative Loop}

  \item \textbf{Action Phase}
  \begin{itemize}
    \item \textbf{Action Module Generation}\\
    The model comprehensively interprets the user's feedback from the previous decision phase (or initial context injection) and the results of the previous action module to generate an exploration or exploitation action module optimized for the current subgoal (task).
    \item \textbf{Action Module Execution and Results Delivery}\\
    The agent executes the generated action module and delivers the collected data and final results to the model. During this stage, the user waits for the action module to complete.
  \end{itemize}

  \item \textbf{Decision Phase}
  \begin{itemize}
    \item \textbf{Results Delivery and Proactive Suggestions to User}\\
    The model presents the results received from the agent in a user-friendly format, such as natural language, tables, or charts. Simultaneously, considering the progress so far and the user's goal, it proposes potential next action modules as proactive suggestions.
    \item \textbf{Human Feedback Delivery}\\
    The user reviews the presented results and suggestions, judges whether the current results are satisfactory, and determines what the next action module should be, providing this as human feedback. Feedback becomes the key input for the generation of the next action module.
  \end{itemize}

  \stagetag{Iterative Loop End Condition}

  \item \textbf{Satisfaction and Termination}\\
  During the decision phase, If the user judges that the results of the action modules within the current subgoal (task) have met a sufficiently satisfactory standard, they provide feedback indicating their intent to terminate the loop. The framework then moves to the next subgoal (task) or, if all subgoals (tasks) are completed, successfully concludes the entire goal.
\end{enumerate}

This Iterative HITL Interaction structure allows BUAs to perform nonlinear, complex, and unstructured browsing that goes beyond the one-way interaction. Furthermore, it enables users to browse in their desired direction with a sense of control, through an appropriate level of interaction, while reducing the physical and cognitive burden.

\section{Use Cases}
This section illustrates how the proposed framework achieves a user's simple and complex browsing goals through two hypothetical use cases. These cases demonstrate how the framework's core concepts—proactive suggestions, human feedback, exploration and exploitation action modules, and iterative HITL interaction—work together organically to specify a user's ambiguous goal and have a BUA perform web browsing to achieve it.

The range of goals achievable through web browsing is diverse, including purchasing, research and information retrieval, using software or products, entertainment/content consumption, and communication \cite{lut2024we}. This section covers two cases. The first case deals with the simple goal of purchasing a product online, and the second addresses the complex goal of conducting market research and sending the results by email.

\paragraph{Scenario 1: Buying milk}
Alice periodically buys milk from online stores such as Amazon and Walmart. Alice prefers products that are inexpensive and offer fast shipping. Due to health reasons, Alice can only eat fat-free milk. Alice wants to find fat-free milk on each online store and purchase the option that best meets her requirements.

\paragraph{Scenario 2: Conducting browser market research and sending results by email} 
Eve, a venture capitalist, is considering investing in a browser startup. She wants to share the results of her research on the global browser market size and growth rate for the next five years with her colleague, Russ, by Gmail. Eve intends to research the browser market's size and growth rate using an appropriate search engine, summarize the findings, and share the information by sending an email to her colleague Russ by Gmail.

\subsection{Use Case 1: Buying Milk}
The first use case explains the process of achieving a simple goal through the framework.

\subsubsection{Goal Setting and Subgoal (Task) Decomposition}
Alice inputs the goal, \textit{``Buy milk for me.''} The framework decomposes this goal into a subgoal (task). Since it is a simple goal, the subgoal (task) is set to the same as the goal: \textit{``Buying milk.''}

\subsubsection{Initial Context Injection for the First Subgoal (Task)}
To gather initial context through proactive suggestions, the model asks questions, \textit{``Which online store would you like to use?''}, \textit{``Do you have a preferred type of milk?''}, and \textit{``What are your most important criteria when purchasing?''} Alice provides a specific contextual information through human feedback: \textit{``Amazon and Walmart,''} \textit{``fat-free milk,''} and \textit{``price and shipping speed.''} Based on this information, the model generates the first action module: \textit{``Search for low-priced, fast-shipping fat-free milk on Amazon.''}

\subsubsection{Iterative HITL Interaction for the First Subgoal (Task)}
The generated action module is executed by the agent. This process consists of an iterative loop of an action phase and a decision phase until Alice terminates the loop.

\paragraph{First Loop}
In the action phase, the agent performs an exploration action module. It navigates to Amazon website, searches for fat-free milk, sorts the products by price, and accesses product pages to check information including shipping options. The information of several products becomes the results of the action module, which is then passed to the model. For instance, the agent could extract information such as \textit{``AAA fat-free milk, \$10, 1 L, same-day delivery available,''} \textit{``BBB fat-free milk, \$8, 500 ml, same-day delivery available,''} and \textit{``CCC fat-free milk, \$8, 500 ml, next-day delivery available.''} In the decision phase, the model shows these results to Alice and proactively suggests the next action: \textit{``Should the agent also search on Walmart?''} When Alice provides the feedback, \textit{``Yes, search on Walmart too,''} the model generates the second action module: \textit{``Search for low-priced, fast-shipping fat-free milk on Walmart.''}

\paragraph{Second Loop}
In the action phase, the agent performs another exploration action module. It navigates to Walmart website and repeats the search process. It passes the results to the model, for example: \textit{``ABC fat-free milk, \$20, 2x1L, same-day delivery available,''} \textit{``AAA fat-free milk, \$12, 1L, same-day delivery available,''} and \textit{``DEF fat-free milk, \$5, 500ml, next-day delivery available.''} In the decision phase, the model presents the results to Alice and proactively suggests, \textit{``Shall the agent compare the collected information to find the best option?''} or \textit{``Should the agent check the reviews for the products found so far?''} When Alice provides the feedback, \textit{``Compare the products already found,''} the model generates the third action module: \textit{``Compare the products found so far to recommend the one with the lowest price and best shipping option.''}

\paragraph{Third Loop}
In the action phase, the agent performs an exploitation action module. It compares and evaluates the results of the previous action modules. Among the six products from Amazon and Walmart, the cheapest milks are \textit{Amazon AAA milk}, \textit{Walmart ABC milk}, and \textit{Walmart DEF milk}. Of these, those available for same-day delivery are \textit{Amazon AAA milk} and \textit{Walmart ABC milk}. Thus, the agent conveys the two options as results of the action module to the model. In the decision phase, the model presents the results to Alice and suggests next action modules such as \textit{``Purchase AAA fat-free milk from Amazon''} or \textit{``Purchase DEF fat-free milk from Walmart.''} When Alice provides the feedback, \textit{``Purchase AAA fat-free milk from Amazon,''} the model generates the fourth action module: \textit{``Purchase AAA fat-free milk from Amazon.''}

\paragraph{Fourth Loop}
In the action phase, the agent performs an exploration action module to purchase the requested product on the Amazon website. Once the purchase is complete, the agent sends the results, \textit{``The purchase of AAA fat-free milk is complete and is scheduled to be delivered the same day to your address''} to the model.

\subsubsection{Iterative HITL End Condition for the First Subgoal (Task)}
The model delivers the results of the action module to Alice. Although it provides a proactive suggestion, \textit{``The purchase is now complete, so I will end the task,''} it does not terminate the subgoal (task) until explicit feedback from Alice. Alice can either end the task or provide different feedback to generate and execute another action module. Since the purchase is complete and Alice judges that satisfactory results have been achieved, Alice requests the model to end the subgoal (task). Consequently, the HITL interaction terminates and the process moves on to the next subgoal (task). Since there is only one subgoal (task) in this scenario, the goal is also achieved, and the entire workflow is completed.

\subsection{Use Case 2: Conducting Browser Market Research and Sending the Email}
\begin{figure*}[h]
  \centering
  \includegraphics[width=0.9\linewidth]{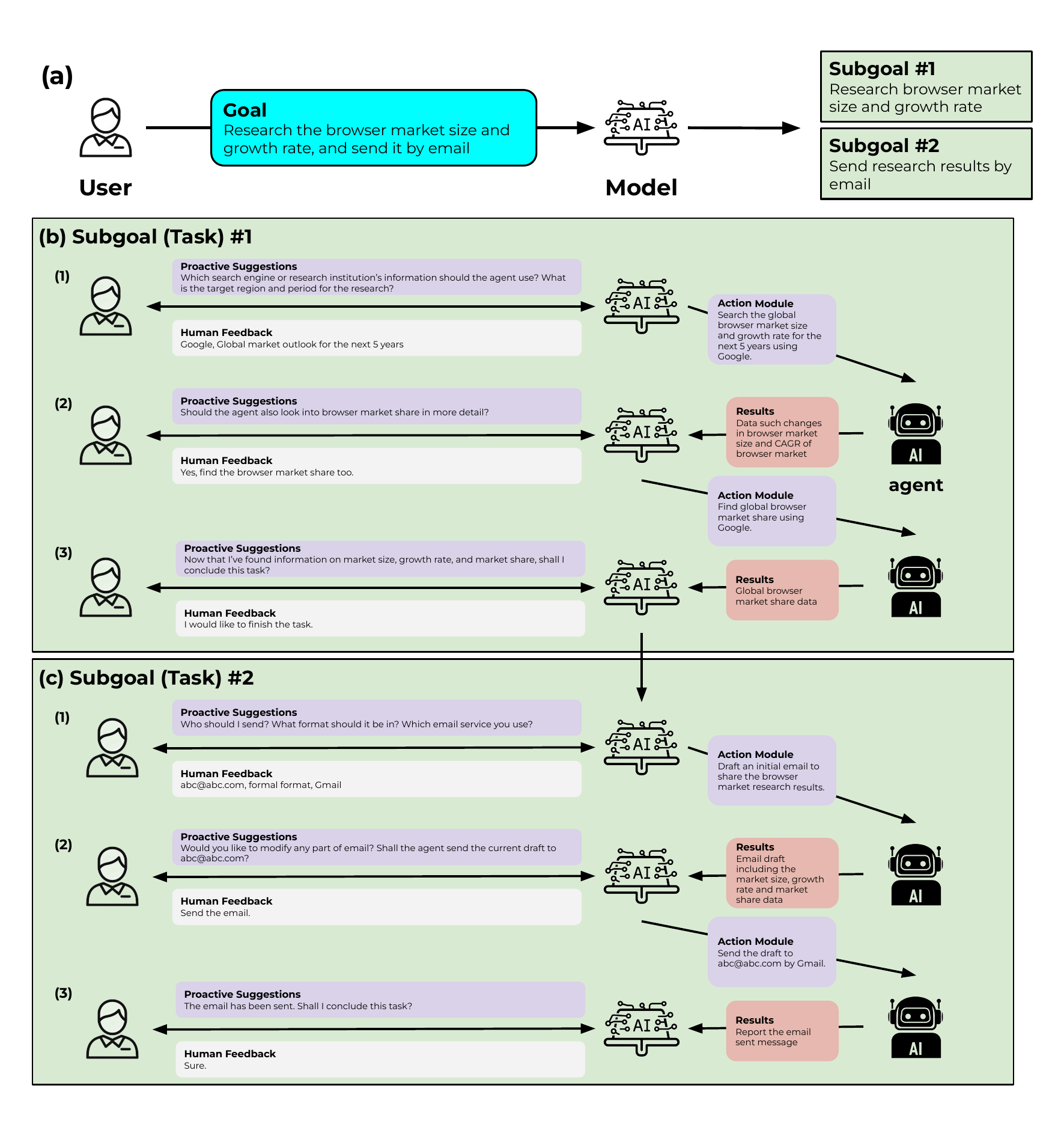}
 \caption{A use case illustrating how the proposed framework handles a complex goal: researching the browser market and sending the results by email. (a) The figure shows the initial goal being decomposed into two subgoals (tasks). (b, c) Within each subgoal (task), it depicts the iterative interaction loop: the process begins with the model's proactive suggestions, followed by the user's feedback, which leads to the generation of an action module executed by the agent, and the results of the action module are then returned to the model and the user. This demonstrates how the framework can support the complex goal of users.}
\Description{
Figure 3 is a storyboard-style diagram illustrating a use case of the framework for the complex goal, "Research the browser market size and growth rate, and send it by email." The figure is split into two main sections for Subgoal #1 and Subgoal #2. Part (b), "Subgoal (Task) #1," shows three turns of interaction between a User, a Model, and an Agent. In turn (1), the Model provides a "Proactive Suggestion" asking for the search engine and research period. The User gives "Human Feedback": "Google, Global market outlook for the next 5 years." This leads the Model to generate an "Action Module" for the Agent to search for this information. In turn (2), the Model presents the "Results" from the first action, then gives a new "Proactive Suggestion" asking about market share. The User's "Human Feedback" is "Yes, find the browser market share too." This generates a new "Action Module" for the Agent. In turn (3), the Model presents the market share data as "Results" and suggests concluding the task. The User agrees. An arrow leads from the end of Subgoal #1 to the beginning of Part (c), "Subgoal (Task) #2," which also shows three turns of interaction. In turn (1), the Model asks for email recipient, format, and service. The User provides the details as feedback, leading to an "Action Module" to draft an email. In turn (2), the Model presents the email draft as "Results" and asks for confirmation to send. The User agrees, generating an "Action Module" to send the email. In turn (3), the Model presents the "Results" confirming the email was sent and asks to conclude the task, which the User confirms.
}
\label{fig:use_case}
\end{figure*}

The second use case explains the process of achieving a complex goal through the framework (Figure~\ref{fig:use_case}).

\subsubsection{Goal Setting and Subgoal (Task) Decomposition}
When Eve enters the complex goal \textit{``Research the browser market size and growth rate of the browser market and send it by email,''} the model decomposes it into two subgoals (tasks). The first subgoal (task) is \textit{``Research browser market size and growth rate,''} and the second is \textit{``Send research results by email.''} After generating the two subgoals (tasks), the iterative HITL interaction begins with the first subgoal (task).

\subsubsection{Initial Context Injection for the First Subgoal (Task)}
To gather context, the model uses proactive suggestions to ask questions, \textit{``Which search engine or research institution's information should the agent use?''} and \textit{``What is the target region and period for the research?''} Eve provides a specific contextual information through feedback: \textit{``Google''} and \textit{``Global market outlook for the next 5 years.''} Based on this context, the model generates the first action module: \textit{``Search the global browser market size and growth rate for the next 5 years using Google.''}

\subsubsection{Iterative HITL Interaction for the First Subgoal (Task)}
\paragraph{First Loop}
In the action phase, the agent performs an exploration action module. It goes to Google and uses various keywords such as \textit{``global browser market size,''} \textit{``global browser market size for the next 5 years,''} and \textit{``global browser market growth rate.''} It accesses multiple links from news and market research institutions, identifies data such as changes in market size and compound annual growth rate (CAGR) and extracts these results. The agent sends these results to the model. In the decision phase, the model shows these results to Eve and proactively suggests \textit{``Should the agent also look at the browser market share in more detail?''} When Eve gives the feedback, \textit{``Yes, find the browser market share too,''} the model generates the second action module: \textit{``Find global browser market share using Google.''}

\paragraph{Second Loop}
In the action phase, the agent performs another exploration action module. It goes to Google, searches using the keyword \textit{``global browser market share,''} and accesses various links such as statistics sites and Wikipedia. The agent finds the current market share by browser and extracts the results. The agent sends these results to the model.

\subsubsection{Iterative HITL End Condition for the First Subgoal (Task)}
The model delivers the results of the action module to Eve. Although it provides a proactive suggestion, \textit{``Now that I have found information on market size, growth rate, and market share, shall I conclude this task?''}, it does not end the subgoal (task) without an explicit request from Eve. Eve can either end the subgoal (task) or provide different feedback to generate and execute another action module. Judging that market research is substantially complete and satisfactory results have been achieved, Eve requests the model to end the subgoal (task). Consequently, the HITL interaction for this subgoal (task) terminates and the process moves on to the next subgoal (task). The information collected is passed on as context for the next subgoal (task).

\subsubsection{Initial Context Injection for the Second Subgoal (Task)}
Now, the framework proceeds with the second subgoal (task). Similarly, the model asks questions such as \textit{``Who should I send it to?''}, \textit{``What format should it be in?''}, and \textit{``Which email service you use?''} Eve provides a specific contextual information through feedback: \textit{``abc@abc.com,''} \textit{``formal format,''} and \textit{``Gmail.''} Based on this information, the model generates the first action module: \textit{``Draft an initial email to share the browser market research results.''}

\subsubsection{Iterative HITL Interaction for the Second Subgoal (Task)}
\paragraph{First Loop}
In the action phase, the agent performs an exploitation action module. It combines the market size, growth rate, and market share data collected in the first subgoal (task) with the context provided by Eve to write an email. This draft is the result of the action module and is sent to the model. In the decision phase, the model presents the result with suggestions \textit{ ``Would you like to modify any part of the email?''} or ``Will the agent send the current draft to abc@abc.com?'' Eve provides feedback, \textit{``Send the email,''} and the model generates the second action module: \textit{``Send the draft to abc@abc.com by Gmail.''}

\paragraph{Second Loop}
In the action phase, the agent logs into Gmail, sets the recipient to \textit{abc@abc.com}, fills in the subject and body with the drafted email content, and sends it. Once the email is sent, the agent reports the results to the model.

\subsubsection{Iterative HITL End Condition for the Second Subgoal (Task)}
The model delivers the results of the action module to Eve. It provides a proactive suggestion, \textit{``The email has been sent. Will I complete this task?''}, but does not end the subgoal (task) without an explicit request. Judging that the email has been sent successfully, Eve requests the model to end the subgoal (task). Consequently, the HITL interaction terminates and the process moves on to the next subgoal (task). Since there are no further subgoals (tasks), the workflow is completed.

\section{Discussion}
In this section, we summarize our proposed HITL interaction framework and its contributions. We then discuss the practical and theoretical implications of the framework and present its limitations and future work.

\subsection{From Browsing to Interacting: Summary and Contributions of the Framework}
The framework suggests the potential for a paradigm shift from traditional web browsing to \textbf{interaction-driven browsing} with BUAs.

\subsubsection{Rethinking the Browser-Using Agent as a Browsing Companion}

The core contribution of this research is the proposal of a conceptual framework that overcomes the limitations of existing BUAs to support complex and nonlinear web browsing processes of users. Conventional BUAs have primarily operated at the level of RPA, characterized as \textit{`single-loop RPA-style linear browsing agents'} designed to execute a clearly defined single goal \cite{zheng2024gpt, gur2023understanding, furuta2024multimodal, abuelsaad2024agent, iong2024openwebagent, lai2024autowebglm, he2024webvoyager}. Although they are effective for repetitive tasks with clear objectives, this approach falls short of supporting the dynamic nature of web browsing, which involves changing goals and adaptive decision-making in various sources of information \cite{pirolli1995information, pirolli1999information, pirolli2003snif, pirolli2005sensemaking, white2009exploratory, marchionini2006exploratory, bates1989design}.

Our framework transforms BUAs into \textit{`iterative loop-based nonlinear browsing agents'} that continuously interact with users to collaboratively specify goals and support the decision-making process. This aims to position BUAs not only as automation or delegation tools, but also as \textbf{browsing companions} that accompany the user throughout the browsing journey \cite{jang2020healthier}. The concepts of the iterative loop and nonlinear browsing are central to enabling this transformation.

The iterative loop refers to the continuous interaction between the BUA and the user. When the user provides feedback, the BUA learns from it to refine and propose subsequent actions. Through this iterative loop, the BUA progressively understands the user's intent, provides a personalized browsing experience, and assists the user in effectively constructing knowledge. Nonlinear browsing means that the BUA is not constrained by its initial plan or specific rules, responding flexibly to user feedback during the browsing process. Our framework decomposes complex browsing tasks into several concrete action modules. Each module represents a linear task that is easy for the BUA to execute. By controlling the direction of each module, the user can lead an overall nonlinear browsing process.

Consequently, this research suggests that a BUA can serve as a companion in the user's complex cognitive processes and nonlinear information exploration journey. This framework is expected to contribute to the design that could improve the efficiency of web-based tasks and browsing.

\subsubsection{Reflecting Human Browsing Behavior}
Another key contribution of this framework is the integration of theories that explain human web browsing behavior into interaction design with BUAs. We have adopted theories such as IFT, satisficing, and the exploration-exploitation trade-off not only as a background but also as core principles of user-centric interaction. This allows users to maintain a mental model similar to their existing browsing experience, even when interacting with a BUA.

This design seeks to distribute the physical and cognitive load that occurs during web browsing between the user and the BUA. Users are freed from repetitive physical operations, such as numerous link clicks and tab switching, allowing them to concentrate on high-level decision-making, such as evaluating the value of information collected by the BUA and determining the next course of action. The BUA assumes the cost of information foraging, while the user concentrates on the critical decisions of assessing the value of the information, aligning the direction of browsing, and determining subsequent actions.

In particular, the loop that continuously generates new action modules by distinguishing between exploration and exploitation action modules is a key mechanism that empowers the user to take the lead in browsing. Web browsing is a continuous process of choosing whether to search for more information (exploration) or to draw a conclusion based on the information collected so far (exploitation) \cite{wiebringhaus1measuring}. Our framework brings this choice process to the forefront of the interaction. Through feedback, users can clearly instruct the BUA on their desired strategy, exploration, or exploitation, thus actively controlling the breadth and depth of their browsing. This directly reflects the natural problem-solving process in which users continue browsing until they reach their satisfactory threshold.

In conclusion, this framework, through an interaction design that reflects human browsing behavior patterns, demonstrates the potential for BUAs to reduce physical and cognitive costs while maintaining the user's mental model of the traditional browsing process. 

\subsubsection{Resolving the Gulf of Envisioning}
Our framework proposes specific mechanisms to resolve the gulf of envisioning, an interaction problem between users and BUAs. In particular, in tasks such as browsing, where the range of possible actions is vast, users experience the gulf of envisioning because they may have an ambiguous goal, such as \textit{``research the market size,''} but struggle to conceive the specific steps required to achieve it \cite{subramonyam2024bridging}.

The proactive suggestions and context injection mechanisms aim to bridge the gulf of envisioning and alleviate the cognitive load associated with user decision-making. Proactive suggestions are mechanisms in which the BUA initiates interaction by asking questions or suggesting plausible next action modules to help specify the user's ambiguous goal. This acts as cognitive scaffolding, informing the user of the possible scope of actions and helping to transform a vague objective into a concrete plan. Users are relieved of the burden of having to formulate a detailed plan; they can clearly convey their intentions by selecting or modifying the suggested options. Context injection is a mechanism that incorporates the user's implicit context, such as preferences, past experiences, and specific constraints, into tasks and action modules. For instance, feedback such as \textit{``I primarily use Amazon and Walmart.''} or \textit{``Price and delivery speed are the most important.''} allows the BUA to align its action modules with the user's personal preferences.

Proactive suggestions help turn a goal into a concrete plan, and context injection ensures that the plan is executed in line with the user's intent. These two mechanisms work in a complementary manner, supporting the user in seamlessly translating an ambiguous goal into the BUA's concrete actions. Through this, our framework contributes to establishing an interaction model in which the BUA first understands and helps specify the user's intent, thereby achieving browsing goals.

\subsection{Design and Theoretical Implications}

\subsubsection{Design Implications for Interacting with Browser-Using Agent}
Our framework presents several implications for the interaction and interface design of BUAs.

\paragraph{Iterative Loop Design Based on User Feedback}
Interacting with a BUA should be designed around an iterative loop that progressively specifies goals through continuous interaction with a user, moving beyond a single command-response structure. Since a user's initial goal is often ambiguous, it must be specified through iterative interactions to deliver a collaborative browsing experience. The interface must clearly visualize the results of action modules within the loop, allow the user to provide intuitive feedback, and transparently show how that feedback is reflected in the next loop.

\paragraph{Explicit Provision of Exploration and Exploitation Action Modules}
To enable users to control their browsing direction and strategy, the exploration and exploitation action modules should be explicitly distinguished and provided. This allows users to intuitively differentiate between searching for more information and deriving results from collected information. Therefore, the interface should either explicitly provide buttons for options of the action modules or clearly transition between exploration and exploitation by discerning the user's intent from their natural language feedback. This is a key design feature that helps users retain control over their browsing.

\paragraph{The Role of Proactive Suggestions}
The BUA should be an active agent that assists user decision-making by proactively suggesting next actions, rather than a passive entity awaiting instructions. In our framework, proactive suggestions serve this role. The interface should explain the context and results upon which the BUA's suggestions are based and present various options to facilitate specific user feedback. This encourages users to provide more detailed feedback, even if they reject the BUA's suggestions. As a result, this reduces the cognitive load of users and provides an experience in which complex browsing is performed step by step through interaction with the BUA.

\subsubsection{Theoretical Implications for Human-AI Interaction}
Our framework has theoretical implications that extend existing theories and offer new perspectives on human-AI interaction, particularly concerning information seeking and decision-making.

\paragraph{Expansion of the Human-AI Collaboration Model}
This framework extends human-AI collaboration beyond simple supervision or delegation to a model of shared agency \cite{cila2022designing,bratman2013shared, wu2025negotiating}. In many existing AI systems, the role of human is focused on reviewing and modifying AI-generated output. In our framework, however, humans and AI are equal partners who divide roles according to their respective strengths. The human leads the strategic direction (what and why to browse), while the AI is responsible for tactical execution (how to browse). This allows humans to focus their cognitive resources on higher-level decision-making, improving collaborative efficiency.

\paragraph{Suggesting the Potential to Overcome Bounded Rationality}
Existing theories such as IFT and bounded rationality hold that humans often stop browsing at a good enough rather than an optimal point due to time and effot costs and physical and cognitive limitations. Our framework suggests a mechanism for reshaping this cost structure by enabling a BUA to lower information-foraging costs. By reducing effective search and evaluation costs, the framework may enable users to pursue browsing paths they would otherwise have forgone, thereby raising satisfactory thresholds and expanding the feasible choice set under bounded rationality.

\section{Limitations and Future Work}

\subsection{Limitations}

As a conceptual framework, this research has several limitations. The primary limitation of this research is the absence of system implementation and user study. Therefore, the real-world efficacy of the framework, particularly whether the iterative loop and the distinction between exploration and exploitation action modules can reduce cognitive load while improving task efficiency, remains a critical open question. Furthermore, the practical success of the framework is highly dependent on the performance of the underlying language models. The quality of the user experience will be determined by the model's ability to interpret ambiguous feedback, generate contextually appropriate suggestions, and reliably execute actions. The potential for poor model performance, which could foster user distrust and confusion, represents a significant technical hurdle for system implementation based on the framework. Finally, contrary to the goal of reducing the cognitive load of traditional browsing, the framework introduces a risk of a new burden, interaction fatigue, by requiring user intervention at each step. Thus, finding the optimal balance between the benefits of user control and the cost of frequent intervention is a crucial challenge that must be addressed in future research and practical implementation.

\subsection{Future Work}
To address the limitations and validate the framework, we propose the following future work. First, a prototype based on the proposed framework should be implemented, followed by a user study. For example, the framework can be evaluated by comparing three conditions: traditional manual browsing, a fully autonomous BUA, and a BUA based on this framework. Furthermore, we can measure several metrics, including task completion time, success rates, perceived cognitive load, and user satisfaction, to assess the efficacy of the framework.

Second, to mitigate interaction fatigue, the framework could be enhanced to automatically enrich the user context by using the user's historical data \cite{cai2025large}. If the BUA can predict user intent and context by analyzing previous browsing history, preferred websites, and repetitive behaviors, it can regulate the level of intervention based on user expertise, predicted intentions, and task complexity. This mechanism is likely to reduce the burden on the user to provide explicit feedback, thus lowering interaction fatigue. 

Third, the framework can be extended from simply summarizing the results of action modules to providing individually customized generative or malleable user interfaces \cite{jiang2025orcabrowsingscaleuserdriven, min2025feedforward, min2025malleable, cao2025generative, chen2025generative}. This would enable the BUA to reconfigure information scattered across multiple websites and visualize it for the user in the most effective format, offering a novel web browsing experience. For example, if the BUA collects products from different websites, the generative user interfaces could display the products from various sources on a unified interface.

Finally, considering that web browsing is not always goal-oriented \cite{lut2024we}, further research is needed on how to seamlessly integrate the \textbf{interaction-driven browsing} of the framework with traditional direct browsing. A hybrid model that allows users to flexibly switch between the two modes depending on the task or personal preference could provide a more comprehensive and satisfying browsing experience.

\section{Conclusion}
In this research, we addressed the limitations of conventional BUAs as single-loop automation tools and proposed a conceptual framework for HITL interaction with BUAs to support the complex and nonlinear web browsing processes of users. Grounded in established theories from cognitive science and HCI on information seeking and human web browsing behavior, the proposed framework is designed to reflect web browsing behavior of users. The framework is centered on an iterative loop that progressively specifies goals, tasks, action modules, and actions, starting from a user's ambiguous goal and evolving through feedback. It is structured around the interplay of the model's proactive suggestions and the user's context injection and decision-making. Furthermore, by defining an agent's sequences of actions as action modules and categorizing them into exploration and exploitation action modules, the framework empowers users to actively steer the agent's browsing direction within the iterative loop until satisfactory outcomes are achieved. Through two hypothetical usage scenarios, this paper demonstrates how the framework can support users' complex and nonlinear browsing processes. Ultimately, this paper presents a vision and the potential for users to achieve their web-based browsing goals by \textbf{interaction-driven browsing} with a BUA rather than browsing directly. Although the framework remains a conceptual model at this stage, we believe that this approach represents a first step toward enhancing the practicality, real-world applicability, and overall user experience of BUAs.

\bibliographystyle{ACM-Reference-Format}
\bibliography{references}


\end{document}